\documentstyle[12pt]{article}
\begin{document}
\author{F. Despa \\
International Centre for Theoretical Physics, \\
PO 586, Trieste, Italy}
\title{On the Imbalance of the Dielectric Constant of the
$PbLi_xNb_{3x}Zr_{0.51}Ti_{0.49-4x}O_3$ Compound}
\date{July, 5 1996}
\maketitle

\begin{abstract}
In the present paper, the dielectric behaviour of the
$PbLi_xNb_{3x}Zr_{0.51}Ti_{0.49-4x}O_3\;\left( PLNZT\right)$ system
is analyzed. It is shown that
the decrease in the dielectric constant with respect to an increase of the 
impurity content may be due to an aggregation trend of the interacting
off-centre dipoles rather than due to cooperative phenomena
which renormalizes the dipole moment of the off-centre impurity.
\vspace{1cm}
\newline
PACS 64.70Kb - Solid-solid transitions \newline
PACS 77.22Gm - Dielectric loss and relaxation \newline
PACS 77.84Dy - Niobates, titanates, tantalates, PZT ceramics, etc.
\end{abstract}

\newpage
Off-centre impurity doped perovskite compunds like 
$PbMg_{1/3}Nb_{2/3}O_3$ $\left( PMN\right) $, $
PbSc_{1/2}Ta_{1/2}O_3$ $\left( PST\right) $, 
$K_{1-x}Li_xTaO_3$, with $x<0.022$, $\left(
KLT\right)$ belong to the large family of relaxors.
These relaxor systems currently receive a great deal of 
interest\cite{1,2,3,4,5,6,7} due to the fact that, in very small concentrations, 
the off-centre impurities are
able to stabilize the ferroelectricity within relatively large domains of the
crystals. Experimentally, for a
relaxor behaviour, the dielectric constant must show a
maximum as a function of the off-centre impurity content\cite{1,2}. When
the off-centre impurity content is increased
the dielectric constant increases initially and then, 
above a critical concentration, it decreases asymptotically. 
The critical concentration above which the dielectric constant
decreases with respect to the impurity content 
is larger than the limiting concentration needed for
''ordering'' to occur. In the case of the $KLT$ system, for example, the 
critical concentration is $x_c = 2.8\times 10^{-2}$\cite{1} while,
relatively large macroscopic, uniformly polarized regions with a length 
scale of at least $10^3-10^4\;\AA $ appears when 
$0.6\times 10^{-2}<x<1.6\times 10^{-2}$\cite{1}. 

Recent experimental results\cite{8} have revealed a similarity between the
dielectric constant of $K_{1-x}Li_xTaO_3$ ($KLT$) and
\newline $PbLi_xNb_{3x}Zr_{0.51}Ti_{0.49-4x}O_3$ ($PLNZT$).
This suggests the existence of a possible off-centre
configuration for the impurities which have been introduced 
in the $PZT$ host matrix.
The behaviour of the
dielectric constant of $PLNZT$ with respect to the impurity
concentration ($Nb$ and $Li$) 
shows a maximum at $x_c = 1.3\times 10^{-2}$. The initial increase of the 
dielectric constant is understood in terms of a self-consistent 
mean-field approach and the theoretical results agree
well with the experimental data\cite{8} (see Fig. 1). The decreasing trend
of the dielectric constant for $x > x_c$ has been tentatively 
attributed\cite{8} to cooperative phenomena which
reduce the effective dipole moment of the off-centre impurity 
(see the curve $\varepsilon _1$ in the Fig. 1).
Agreement with experimental data is rather poor.

As is known, the most intriguing property of off-centre systems is
that they carry a permanent electric dipole moment. Because of this,
when impurities of this kind are introduced in a host matrix,
the permittivity $\varepsilon $
of the crystal increases and becomes strongly dependent on frequency\cite{2}.
Increasing the content of the off-centre dipoles within a well-established
range, the
ferroelectric interaction strengths between off-centre dipoles 
overcomes the strength
of the extra random fields. This  stabilizes the
ferroelectricity\cite{2,5}. Moreover, when the dipole
concentration exceeds the critical one (usually, larger than 
the limiting concentration needed for ''ordering'' to occur\cite{1}),
additional complexity in the dynamical
behavior may appear. On one side, dipolar correlations occur, and these not only
reduce the polarizability of the system\cite{2} (this decreases the dielectric
constant) but functionally change
the dipole-dipole interaction energy dependence on the spacing of 
dipoles\cite{2,5}. On the other hand,
the dipolar interaction between randomly distributed off-centre impurities
facilitates the aggregation of a small fraction of impurity content. 
See, for example, the case of the $KLT$ system;
$(Li^{+})_2$ are most frequently present above the critical 
concentration\cite{3,9,10,11}. This
leads to an extremely long relaxation time\cite{9,11,12}, having
implications for the hysteresis loop as well as for the dielectric 
constant\cite{5}. 
The same behavior has been qualitatively seen within the $SCT$
system\cite{7} where, in addition to conventional non-linear response, 
important paraelectric cluster contributions were revealed. As in the previous
case, this may be understood as a consequence of the aggregation of
the interacting off-centre $Ca^{++}$ dipoles. 

In the present paper, it is shown that
the decrease in the dielectric constant of the $PLNZT$ system 
with respect to the increase in 
impurity content\cite{8} may be due to the aggregation 
trend of the interacting
off-centre dipoles rather than due to cooperative phenomena which
renormalize the dipole moment of the off-centre impurity.

The initial increase of the dielectric constant with respect to impurity 
content (see in Fig. 1 below of $x_c$) can be described by a
self-consistent mean-field theory using a Clausius-Mossotti equation 
\begin{equation}
\label{1}\frac{\varepsilon -1}{\varepsilon +2}=\frac{\varepsilon _0-1}{%
\varepsilon _0+2}+\frac{4\pi }3x\left( \chi _{o\;Li}+3\chi _{o\;Nb}\right)
\;, 
\end{equation}
where $\varepsilon _0$ is the permittivity of the host matrix which in our
case has a high polarizable character, and
$\chi _{o\;Li,Nb}$
stands for the single-particle polarizability of noninteracting
impurities. In the framework of classical theories,\cite{2} the latter is
expressed by 
\begin{equation}
\label{2}\chi _{0\;Li,Nb}=\frac{2d_{Li,Nb}^2}{3\Delta }th\left( \frac \Delta
{2kT}\right) \simeq \frac{d_{Li,Nb}^2}{3kT}\;, 
\end{equation}
where $\Delta $ is the tunneling energy and, usually, $\frac \Delta {k_B}%
\simeq 1K$, and $d_{Li,Nb}$ are the intrinsic dipole moments determined by
the off-center displacement magnitude of $Li$ and $Nb$ respectively. 
Agreement between the theoretical (eq. 1) and experimental curves for the
increasing region was obtained for 
$d_{Li}^2+3d_{Nb}^2\simeq 0.0075\;e^210^{-20}m^2$\cite{8}. 
When cooperative phenomena occur, the single-particle polarizability
is expressed by\cite{2}
\begin{equation}
\label{3}\chi _{Li,Nb}^{\circ }=\chi _{o\;Li,Nb}\left( 1-x\frac{4\pi }3\frac{%
d_{Li,Nb}^{\circ 2}}{kT\varepsilon _0}\right) \;,
\end{equation}
where 
\begin{equation}
\label{4}d^{\circ }=\frac{\gamma \varepsilon _0}3d_{Li,Nb}
\end{equation}
is the effective dipole moment of the impurity in a highly polarizable
matrix\cite{2}.
Equation $\left( 3\right) $
correctly describes the single-particle polarizability within the 
framework of the
self-consistent mean-field approximation which means $x\gg x_c$. In this
case $\left( x\gg x_c\right)$, the mean spacing between the impurities
becomes small enough to reduce the fluctuations of the local fields (it is
known that the existence of the fluctuations of the local fields 
invalidates the mean-field theories\cite{2}). The dielectric constant for
the decreasing part has been calculated\cite{8} using (3) (see in Fig. 1
the curve $\varepsilon _1$). As one can see
agreement with the experimental data in this region is rather poor.

Let us assume that, for different reasons (mainly, related to
the off-centre dipole pairwise interaction discussed in the 
introduction), the impurities aggregate in small clusters.
In all probabilities, the
aggregation process is strongly mediated by the peculiar
nature of the off-centre
crystalline field which, as is known, has a certain number of separate
potential minima around the normal position in the host lattice.
Another argument which supports the above supposition is the case
of Na-doped
fullerites\cite{13} where the solute is identified both in off-centre
locations\cite{14,15} and in small clusters within the octahedral
interstices\cite{16}. In our case,
the aggregation phenomenon may lead to the appearance of an effective dipole
moment (smaller than that corresponding to a free off-centre dipole).
On the other hand, it may obstruct, 
progressively, the reorientation possibility 
of the permanent off-centre dipoles until these are complete blocked.
The former case occurs as long as the potential wells of the off-centre 
configuration is only partially occupied. The latter case occurs
when the dipoles are completely blocked and should coincide with the
occupation of all the potential wells of the off-centre configuration.
In fact, concerning the last situation, we may say that 
the dipole moment no longer exist. Note
that, actually, the off-centre potential is realized for a single
substitutional particle interacting with the host ions. The balance
between the repulsive and the polarization forces destabilizes the central
equilibrium. In the case of more particles, the
potential curvature of the off-centre configuration may be drastically
changed, but it will be disregarded herein.

The aggregation phenomena assume that in our case, over the critical
concentration, the substitutional impurities are not uniformly
spread over the corresponding vacancy sites of the host matrix. They prefer,
with a higher probability, to form small intercalated clusters (dimers,
trimers, etc). In this view, the added impurities, whose number is
$\left(N - N_c\right) $, where $N_c$ is the initial (critical) number of 
off-centre impurities, get places on the available off-centre sites which are
in a total amount of $n \times N_c$, where $n$ is the number of the 
potential wells in a given
off-centre configuration. The remaining number of the free off-centre dipoles
is $N_c \times (1 - P) - (N - N_c) \times P = N_c - N \times P$, where $P$
is the probability for an added impurity to occupy one potential well 
of the off-centre configurations. Neglecting the fact that this probability
should depend on the strength of the off-centre dipole pairwise interaction,
it may be readily approximated by
$ P = \frac{(nN_c - (N - N_c))!}{nN_c ! (N - N_c)!}$. The
dielectric constant of the system characterized by the above number
of the off-centre dipoles becomes  
\begin{equation}
\label{5}\varepsilon _2=\frac{\frac{\varepsilon_o -1}{\varepsilon_o +2} 
+2\left( x_c - Px\right) \chi _{o\;Li,Nb}}
{\frac{\varepsilon_o -1}{\varepsilon_o +2} 
- \left( x_c - Px\right) \chi _{o\;Li,Nb}}\;.
\end{equation}
In Fig. 1 experimental data is shown for the dielectric constant of
the $PLNZT$ system versus the impurity content\cite{8}. Also, theoretical
curves are given for its decreasing region which correspond both to the cooperative phenomena 
approch (see $\varepsilon _1$) and to the aggregation phenomenon approach 
(see $\varepsilon _2$).
As one can observe in the case of the curve $\varepsilon _2$ 
agreement with the experimental data is improved. This means that for the
$PLNZT$ system, there exists an aggregation trend of the off-centre impurities 
when their content overcomes a critical point. 

Within the present paper, we have shown that, under some circumstances, 
especially related to the aggregation phenomenon of impurities
mediated by off-centre configurations, the dielectric constant corresponding to
the off-centre impurity doped crystals may decrease with the
increase of the off-centre
impurity content. Aggregation phenomena of these kinds of impurities 
have been seen
experimentally by various techniques \cite{7,9,12,16} and, as one observes,
studying the change of the dielectric
constant by the increase of the content of the off-centre 
impurities may also, be a useful tool in this respect. 
\newline
{\it Acknowledgement.} The author thanks Dr. J. Lawson for
technical help. Also, he would like to thank Professor E. Tosatti for his
kindly hospitality during the stay at ICTP - Trieste.

\newpage

{\it Figure Captions}

\vspace{3cm}
\noindent
Figure 1  - Dielectric constant $\varepsilon $ versus impurity content 
$x$ for the $PLNZT$ system (experimental data available from \cite{8};
 $\varepsilon _1 $ - theoretical results within the cooperative phenomena 
approach, $\varepsilon _2$ - theoretical results within the aggregation 
phenomenon approach.

\end{document}